\documentclass[pre,twocolumn,showpacs,superscriptaddress,amsmath] {revtex4}
\usepackage{graphics}

\newcommand{\cu}
{\affiliation{Department of Physics, University of Calcutta, 
92 Acharya Prafulla Chandra Road, Kolkata 700009, India}}

\begin{document}

\title
{Complex Networks: effect of subtle changes in nature of randomness}

\author{Sanchari Goswami} 
\cu
\author{Soham Biswas} 
\cu

\author{Parongama Sen}
\cu


\begin{abstract}

In two different classes of network models, namely, the Watts Strogatz type and the
Euclidean type, subtle changes have been introduced in the randomness. In the Watts Strogatz type network,
rewiring has been done in different ways and although the qualitative results remain same, 
finite differences in the exponents are observed. In the Euclidean type networks, where at least 
one finite phase transition occurs,
two models differing in a similar way have been considered. The results show a possible shift in 
one of the phase transition points but no change in the 
values of the exponents. The WS and Euclidean type models are equivalent 
for extreme values of the parameters; we compare their behaviour for intermediate values. 
\end{abstract}

\maketitle

\section{Introduction}

Ever since the discovery of small world effect in many natural
and artificial networks, many models have been proposed to mimic their features
\cite{bareview}. Basic properties like the small world effect can be obtained from the Watts Strogatz (WS) model \cite{WS} while 
a prototype of the scale-free model is
the Barabasi Albert model \cite{BA}. For networks in which links depend on the geographical distance or age separating the nodes, a few models have also
been proposed \cite{psreview,doro}. 

In all these models, an essential ingredient is randomness. Randomness
is involved in the way the nodes are connected in the network. For example,
in the original WS model, links between nearest neighbours were rewired 
with a certain probability to a randomly chosen node.

While randomness is the key feature, one may ask the question, to what extent
does its nature affect the network? 
For example, one can have a  
variation of the WS  model, in which links are added randomly keeping the original
links intact \cite{newman}.  Thus one can conceive of  
 two kinds of WS models, the rewiring type and the addition type.
The network properties are in general identical for the two types.

Another class of models which has been studied extensively includes the evolving networks. In these networks, 
a new node prefers to get  attached to older nodes  characterised by some particular feature like  
 degree,  age or geographical proximity     etc.
In an Euclidean network, links are established between nodes with a
probability which depends on the Euclidean distance separating them.
These probabilities   can  be tuned to change the nature of the
randomness.  
Changing the degree of the preferential attachment in an
evolving network has been shown to lead to 
drastic changes its behaviour \cite{redner}. However, this variation
is much more severe compared to the example of  addition and rewiring type WS models.

In this paper, we have considered two classes of models. One is the WS type
with rewiring and the other class is Euclidean networks in which links
are attached with the probability  proportional to $l^{-\alpha}$ where $l$ is the
Euclidean distance separating them.
 Within  both classes, we now
introduce subtle changes in the randomness (to be elaborated in the 
appropriate sections) and see how it affects the (static) network properties. 
At one level, we discuss the effect  induced by the change in randomness
within each class and on a different level, we compare the WS and the Euclidean networks which  are equivalent for extreme values of the parameters.

In section II, we briefly describe the models and the network
properties which have been studied. 
The WS type networks have been discussed  in section III
and in section IV, the Euclidean network results have been presented. Comments and 
discussions are made in Section V.

\section{The  models and the network properties studied}

 To construct the  WS type of networks, we start  with  a regular network 
with $2K =4$ nearest neighbours. 
In the first type of WS network, henceforth referred to 
as WSA, we rewind both the first and second neighbours 
with probability $p$. In the other type, {\it{only}} the second 
neighbours are rewired and we call it the 
WSB.

 WSB may be looked upon as a particular case
of  WS networks with $2K$  original links for each node and where 
rewiring of the 2nd, 3rd....$K$th links are done only, i.e., the first nearest neighbour links are kept intact.
Such a case was recently considered with rather
high values of $K$ and the results were found to be identical with WS \cite{new}. This result is not surprising, as the number of rewired links for large $K$ is considerably larger than the number of links which are not rewired. 
In case of $K=2$, i.e., the present case, however, the effect (if any) of this kind of 
partial rewiring on the network properties is expected to be much more prominent   as
the number of links which are being rewired and those which are not,
are both equal to $L$ (the number of nodes),  in model WSB.

\begin{figure} [h]
\rotatebox{0}{\resizebox*{8cm}{!}{\includegraphics{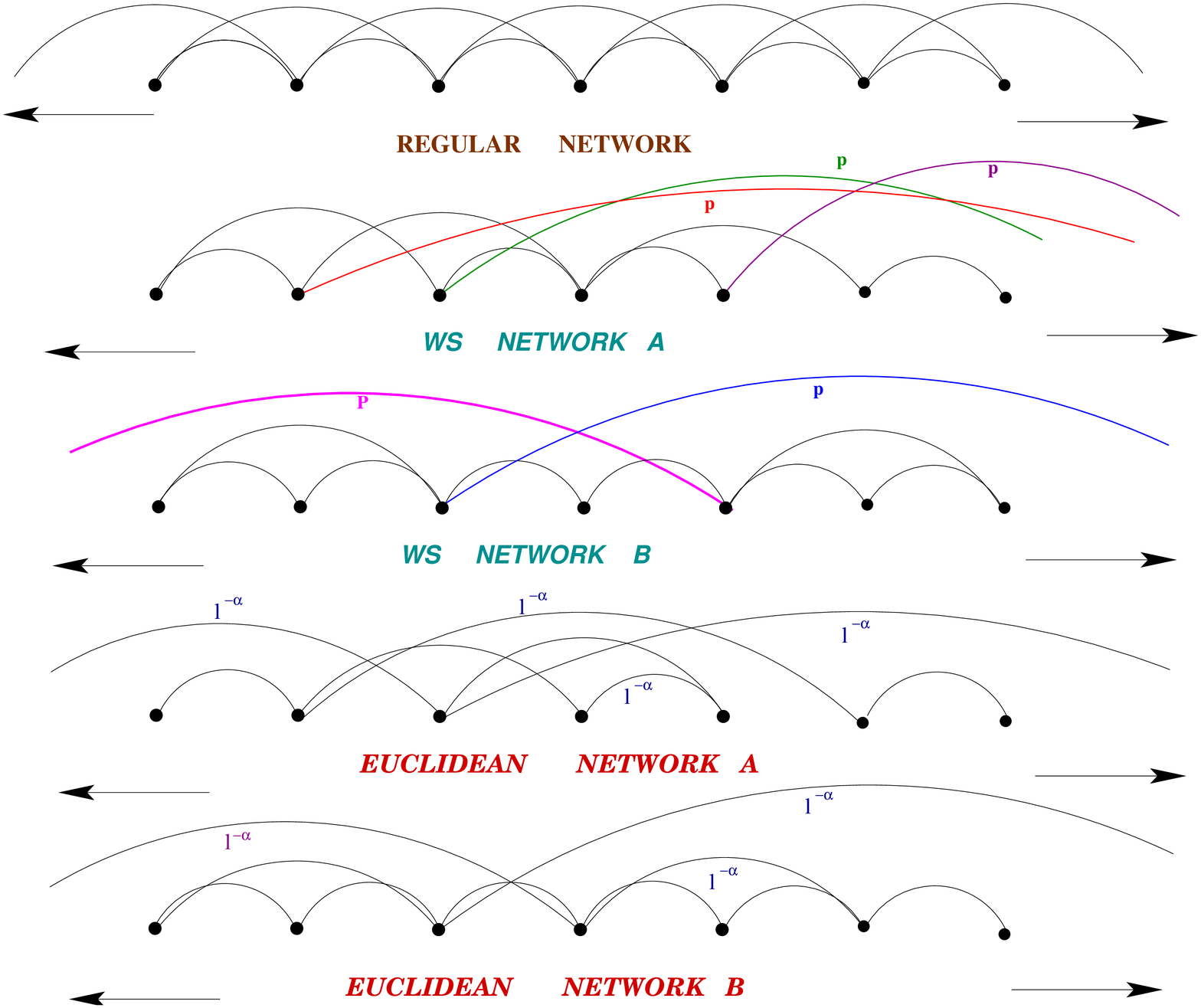}}}
\caption{(Color online) Schematic diagram for different network models. Average
 degree is $2K=4$ in each network. In the regular network both the first and second nearest neighbour are present. In WS network A both the neighbours are rewired while for WS network B only second neighbours are rewired with probability $p$.
In Euclidean network A nodes  at any $l \geq 1$ are linked while in Euclidean network B, the first nearest neighbours are always linked while other nodes are linked with the probability $l^{-\alpha}$ with $l \geq 2$.
}
\label{schm}
\end{figure}

In the  Euclidean network,  nodes placed at a distance $l$ along a one
dimensional chain are linked with a probability $P(l) \propto l^{-\alpha}$. In the first type of model which we  call Euclidean A type (EA),
nodes  at any $l \geq 1$ are linked while in the second type, EB, the first nearest neighbours are always 
linked while other nodes are linked with the probability $P(l)$ with $l \geq 2$.
A schematic picture of the networks is given in Fig. \ref{schm}.

In both WSB and EB, a broken network will never appear.
The Euclidean network of type B was first  studied in the context of navigation \cite{psreview} and later in relation to  polymers \cite{blumen,psbkc,mouk} where
short cuts or bridges appear in a comparable manner. Other studies on the EB were later conducted  
where  issues like clustering properties, navigation with local information and phase transition of the Ising model were   
addressed
\cite{biswas,zhu,psen_arnab,new2,psnetrev}. 
 The   results of the various earlier studies 
indicate that the EB model behaves as a regular network for $\alpha > 2$ and
like a small world for $\alpha \leq 1$. For $1 < \alpha  < 2$, there
is some controversy regarding  the behaviour of the model; while some studies
suggest that it is small world like, there are others which indicate
a regular lattice (of dimension greater than one)-like behaviour \cite{psnetrev}.
In \cite{biswas}, it was claimed that there are two transition 
points and the region $0 < \alpha < 1$ is 
a random graph with vanishing clustering tendency. 
In these  studies, EB has been considered with
two nearest neighbours plus one (on an average) long range  link for each
node. The main motivation for the present study comes from the fact
that EA has never been studied earlier. Here we consider for  EA, 
four random long range links for each node
on an average and for  EB, two nearest neighbour links plus two random 
long range links 
(on an average). This is done  to  make the 
Euclidean networks
comparable  with the Watts Strogatz like network for which the average 
degree is four.

The network properties we have studied are the basic ones like average
shortest paths, clustering coefficient and degree distribution.
The average shortest paths have been calculated {\it{exactly}} using 
a burning algorithm.

The two types  of models, WS and Euclidean,   are equivalent and identical to a random network
 in the limit $p=1$ and $\alpha =0$.
They are also   equivalent   
for $p =0$ and $\alpha = \infty$, corresponding to a 
 regular chain of nodes  with  four nearest neighbour links for each node.
 For this regular network  the average
shortest path is 
\begin{equation}
\langle s \rangle = \dfrac{(L+3-X)(L-1+X)}{8(L-1)},
\label{reg-short}
\end{equation}
where $L$ is the number of nodes or system size and $X$ is the remainder when $(L-1)$ is divided by $4$.

Clustering coefficient of  the $i$th node is defined by  
\[
C_{i}= \frac{T_{i}}{[k_{i}(k_{i}-1)/2]},
\]
where $T_{i}$ is the number  of closed triangles attached with 
the node $i$ and 
$k_{i}$ its degree.
 The average clustering coefficient of the regular network at $p=0$ or
$\alpha = \infty$ is 0.5. 

Degree distribution is another important property of a network.
 Both the WS type and Euclidean networks are homogeneous, which means that
the 
degree distribution will have a typical scale.  For the regular network obtained in the the extreme limits $\alpha \to \infty$ and
$p=0$, the degree distribution 
is simply 
a delta function at $k=4$.

\section{Watts Strogatz type networks}

We have calculated the average shortest paths as a function of the
number of nodes $(L)$ and rewiring probability $(p)$. For both WSA and WSB,
 the average shortest path varies as $\ln(L)$ for $p \neq 0$. 
But when the variation against the rewiring probability $p$ is studied, there appears to be some 
quantitative difference in the results  between WSA and WSB.

It is now widely accepted that the average shortest path obeys the general scaling form \cite{bareview}
\begin{equation}
 \langle s \rangle \sim \dfrac{N^{1/d}}{K}f(KpN),
\label{scws}
\end{equation}
where  $N$ is the total number of nodes in a space of dimension
$d$ and

$~~~~~f(KpN) = \dfrac{\ln(KpN)}{KpN}~~ $  for   $~~KpN >> 1.$

In the present case,   the dimension of the system $d=1$ and $N=L$, such that 
equation (\ref{scws}) becomes

\begin{equation}
 \langle s \rangle \sim  (1/K^{2})\dfrac{\ln(p)+\ln(KL)}{p} ~~~~~{\rm{for}}~~~~ KpL >>1.
\label{scwss}
\end{equation}

\begin{figure} [h]
\rotatebox{0}{\resizebox*{8cm}{!}{\includegraphics{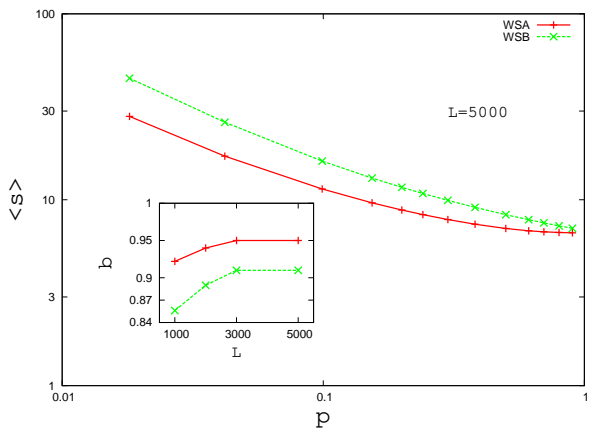}}}
\caption{(Color online) Variation of average shortest path as a function of rewiring probability $p$ for the WSA and WSB models $(L=5000)$. Inset shows the variation of $b$ with system size.
}
\label{avsp}
\end{figure}

However, from the numerical results  for finite systems shown in Fig. \ref{avsp}, we find the following scaling form to be more appropriate
\begin{equation}
 \langle s \rangle \sim a\dfrac{\ln(p)+c}{p^{b}},
\label{scwsab}
\end{equation}
where $b$ is not exactly equal to unity and depends on the system size. The value of $b$, when plotted against
$L$, indicates that it saturates at large values of $L$, with 
the saturation values close to but not exactly equal to unity ( see inset of 
Fig. \ref{avsp}). Moreover, 
the saturation values of $b$ ($\sim0.95$ for WSA and $\sim0.91$ for WSB) have a finite 
difference for the two models at least over the range of values of $L$ considered here.

\begin{figure} [h]
\rotatebox{270}{\resizebox*{5cm}{!}{\includegraphics{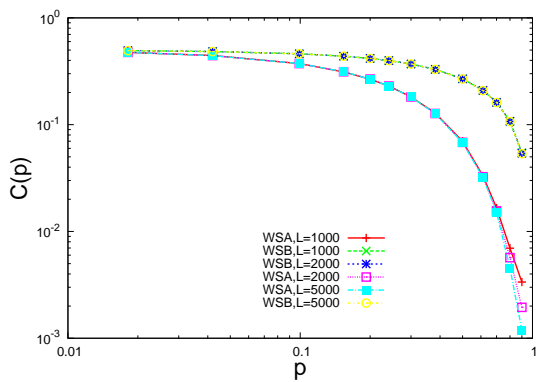}}}
\caption{(Color online) Average clustering coefficient $C(p)$ as a function of 
rewiring probability $p$ for the WSA and WSB model for different system sizes are shown. 
$C(p)$ is fitted to the form $C(p) \sim (1-p)^x$ with $x=3$ (WSA) and $1$ (WSB).
}
\label{ccws}
\end{figure}

The average clustering  coefficient for $p=0$ is 0.5 as mentioned before. 
For $p>0$, two neighbors of a node $i$ that were connected
at $p=0$ are still neighbors of $i$ and connected by an
edge with probability $(1-p)$. Since there are three
edges that need to remain intact for a cluster,  $C(p)\simeq C(0)(1-p)^3$ in 
WSA.
For WSB $C(p)\simeq C(0)(1-p)$ as we do not rewire the first neighbours in this case. This is 
exactly what we observe in the numerical studies (Fig. \ref{ccws}).
Hence WSA and WSB have clearly different exponents as $p \to 1$, the point at which the transition to random network is obtained.

\begin{figure} [h]
\rotatebox{0}{\resizebox*{9cm}{!}{\includegraphics{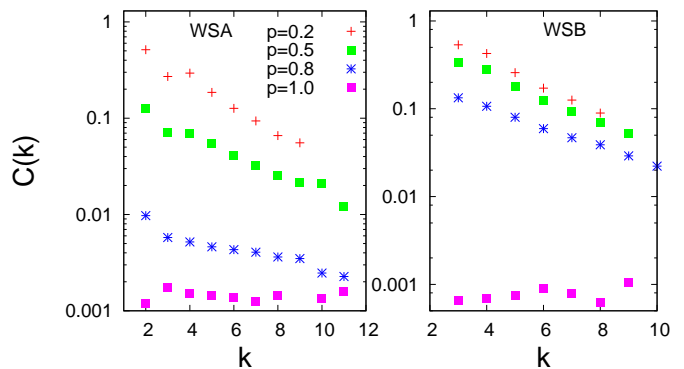}}}
\caption{(Color online) Average clustering co-efficient as a function of the 
degrees ($C(k)$ vs $k$) for different rewiring probabilities 
for the WSA and WSB model for system size $L=2000$.
}
\label{ckwsab}
\end{figure}

The clustering coefficient as a function of the degree is another important quantity in a network. We find that in the WS networks, the maximum degree is 
 about 10, the clustering coefficients show an exponential decay with $k$ within this small range of $k$ (Fig. \ref{ckwsab}). 
For WSA the exponential decay becomes much weaker as $p \to 1$, while for WSB, the slope in a log-linear plot shows a much stronger dependence of $C(k)$ on $k$ even for large value of $p$.
The clustering coefficients rapidly decrease as $p$ increases, as is expected, and at $p=1$ becomes very close to zero showing no dependence 
on $k$.
The clustering coefficients are obviously larger in WSB than in WSA in magnitude (Fig. \ref{ckwsab}) and 
do not show any dependence on the system size for either model.

\begin{figure} [h]
\rotatebox{0}{\resizebox*{8cm}{!}{\includegraphics{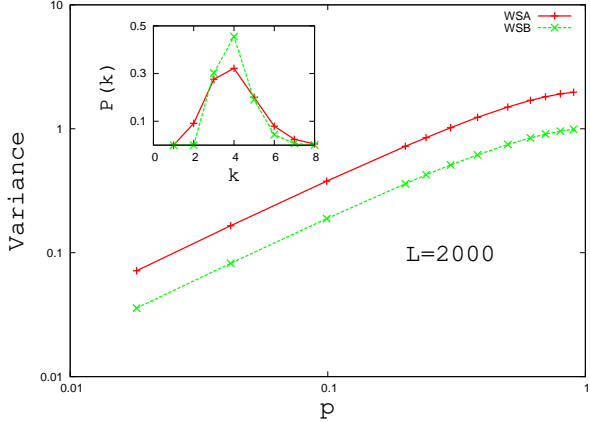}}}
\caption{(Color online)  Variance of the degree distribution as a function 
of rewiring probability $p$ for the WSA and WSB model for system size 
$L=2000$ are shown. Inset shows the degree distribution $P(k)$ against $k$.
}
\label{ddws}
\end{figure}

The degree distribution, which  at $p=0$ is a delta function,  shows a spread as $p$ is made nonzero. The variance of the 
degree distribution as a function of $p$ can be studied. The results show that 
the nature of variance is identical in WSA and WSB, it shows a monotonic  increase
with $p$ as expected [Fig. \ref{ddws}].

One obvious difference between the 
 WSA and WSB  regarding the degree distribution is that $P(k \leq1)=0$ for WSA and 
$P(k \leq 2)=0$ for WSB because of the way the rewiring 
is done \cite{bareview}.  
 On the other hand the other probabilities, e.g., 
$P(k > 2)$ for WSB  are nonzero and almost independent of the system sizes considered.


\section {Euclidean networks}

In the Euclidean network, we have similarly two models
which differ in a subtle manner as described in section II.

\begin{figure}[h]
\rotatebox{0}{\resizebox*{8cm}{!}{\includegraphics{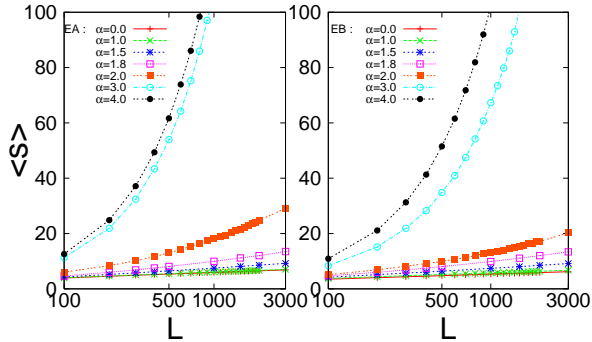}}}
\caption{(Color online) Log-linear plot of variation of average shortest 
path against system size 
for different values of $\alpha$ for EA and EB. The behaviour 
of $\langle s \rangle$ apparently deviates from 
$\langle s\rangle \propto \log L$ even below $\alpha=2.0$.
}
\label{eavsp3}
\end{figure}

In Fig. \ref{eavsp3}, we present the result for the shortest paths against 
$L$ for both EA and EB models for different $\alpha$. 
The log-linear plot shows that there could be a deviation from the behaviour 
$\langle s\rangle \propto \log L $ even below $\alpha = 2.0$. This 
is in consistency with some earlier results \cite{mouk,biswas,psen_arnab} 
which claim that there is another 
phase transition at $\alpha =1$. We have not 
attempted a very detailed study to check exactly for which value of $\alpha$ the small world 
feature (i.e., $\langle s\rangle \propto \log L $) vanishes as one needs huge system sizes 
for that and 
exact numerical calculations becomes difficult.

\begin{figure}
\rotatebox{270}{\resizebox*{5cm}{!}{\includegraphics{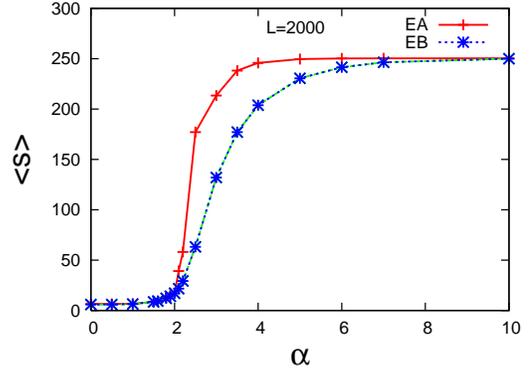}}}
\caption{(Color online) Comparison of average shortest path against $\alpha$
 for EA and EB. In the region 
$2 < \alpha < 10$ we have finite differences in 
$\langle s\rangle$.
}
\label{eavsp2}
\end{figure}

When we plot the shortest paths against $\alpha$,
it is apparent that there is 
a transition at $\alpha = 2$ beyond which both EA and EB behave as regular networks. 
This is shown in Fig. \ref{eavsp2}. This result is well established in all previous studies. When we compare 
the shortest paths for EA and 
EB, we find that finite differences in the 
results appear for $2 < \alpha < 10$ and the 
transition at $\alpha =2$ looks much sharper in EA. 
At large values of $\alpha$, for both EA and EB, the 
shortest paths saturate at a value corresponding to 
that obtained from (\ref{reg-short}) as expected.

\begin{figure}
\rotatebox{0}{\resizebox*{7cm}{!}{\includegraphics{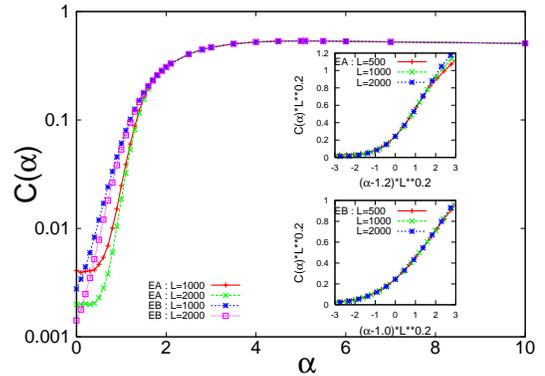}}}
\caption{(Color online) Variation of average clustering coefficients 
with $\alpha$ for  $L=1000$ and $L=2000$ are shown for EA and EB models. Insets show the scaling plots for EA (top) and EB (bottom). 
}
\label{eccki}
\end{figure}

The average clustering coefficients calculated from these networks are
shown as functions of $\alpha$ in Fig \ref{eccki}. 
Once again, the values saturate at large values of $\alpha$ to the
expected value $1/2$.

\begin{figure} [h]
\rotatebox{0}{\resizebox*{5cm}{!}{\includegraphics{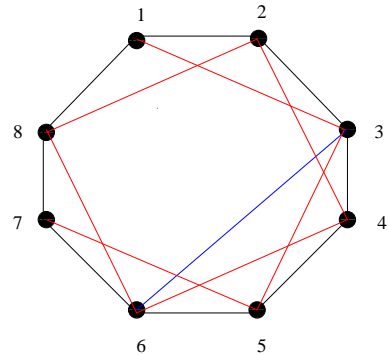}}}
\caption{(Color online) A toy model of EB type in which the average degree is four. The clustering coefficient of this model is $\simeq 0.53$, larger than $0.5$, the clustering coefficient of a regular network.
}
\label{toymodel}
\end{figure}

There are, however, several interesting 
facts to be noted. First, we find that clustering coefficient $C(\alpha)$ is 
marginally larger than $1/2$ for some values of $\alpha > 2$ indicating that 
a network with some randomness can have larger clustering 
coefficients compared to the corresponding regular one.
We present in Fig. \ref{toymodel} a toy model which has larger clustering coefficient
compared to the regular one to show that it can be possible.

In the Euclidean models, the system size dependence of the clustering 
coefficient shows an interesting result. Normally, one does not expect that clustering coefficient will
have size dependence but we do get such a result here in these type of networks. According to \cite{biswas},
it is indeed possible if there is a transition point below which the 
clustering is zero and above which it is finite. Hence, as in \cite{biswas},
we attempt a data collapse by scaling the variables assuming 
$C(\alpha)$ to satisfy the following behaviour
\begin{equation}
C(\alpha) \propto L^{-\gamma} f[(\alpha - \alpha_c)L^\delta],
\end{equation}
where the scaling function is a constant for very small values of the argument 
and 
varies as $[(\alpha - \alpha_c)L^\delta]^{\gamma/\delta}$ in the opposite limit.

Using the above form, we find excellent data collapse 
for different sizes with $\gamma = \delta \simeq 0.2$ for both EA and EB ( see insets 
of Fig. \ref{eccki}). However, the value of  $\alpha _c$ is different; $\alpha_c \simeq 1.2$ for
EA and $\alpha_c \simeq 1$ for EB. The above analysis 
 gives us a lot of important information:
the exponents are independent of EA and EB while the transition points are not.
 The scaling functions
are also different for the two. 
The values of $\gamma$ and $\delta$ being almost same showing that the clustering is 
a linear function of  $\alpha-\alpha_c$ at $\alpha >> \alpha_c$.

\begin{figure}
\rotatebox{0}{\resizebox*{8cm}{!}{\includegraphics{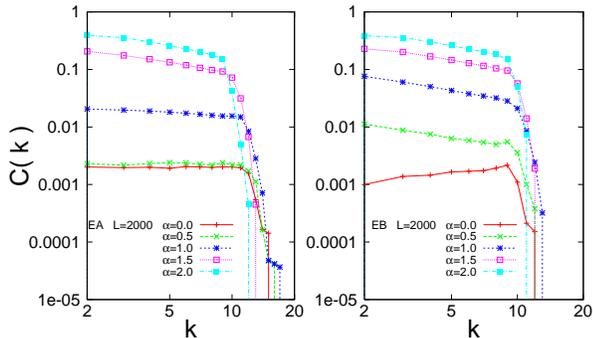}}}
\caption{(Color online) Variation of clustering coefficients against degree for 
EA and EB for $L=1000$.
}
\label{ecck}
\end{figure}

As a function of the degree $k$, the clustering coefficients are plotted
against $k$ in Fig. \ref{ecck}.
The values of $k$ are limited and $C(k)$ shows weak dependence on $k$ and the variations do not show any drastic change with $\alpha$ in either EA or EB models.


\begin{figure}
\rotatebox{0}{\resizebox*{7cm}{!}{\includegraphics{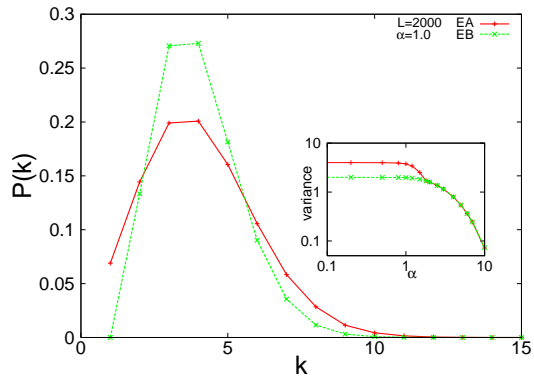}}}
\caption{(Color online) Comparison of degree distribution of 
EA and EB for system size 
$L=2000$.
}
\label{edd}
\end{figure}

Next we consider the degree distributions. For both EA and EB, these are 
shown in Fig. \ref{edd}. For EA, minimum value of the degree
can be zero while in EB, there is a lower cutoff at 2. Consequently,
the degree distributions, both of which are peaked around the average value $4$, have different variances as function of $\alpha$ which are plotted in the inset.  Here we find that the variance for the two models are significantly 
different upto $\alpha \simeq2$;  for EB, it is constant upto $\alpha = 1$ and falls off
smoothly, but for EA, it has a different constant value till $\alpha = 1$, develops a kink between $\alpha = 1$ and $2$ before
converging with the EB values beyond $\alpha = 2$.  

\section{Summary and Discussions}

We considered network models with a fixed average degree $2K=4$ 
where the randomness in linking the nodes has been conceived in different ways.

The WSA and WSB networks, as demonstrated in section III, do not show
qualitative differences in the main  features. Finite differences in some 
exponents are there; e.g., the value of 
$b$ appearing in the shortest path expression shows a difference. It is expected that for WSA the value of $b$ should 
approach unity in the thermodynamic limit. Similarly, 
for WSB too the value may approach unity for infinite systems. However for the
finite sizes considered here, we get different values for WSA and WSB neither of which is equal to unity. Since we evaluate the shortest paths exactly (numerically) it is difficult to study very large 
sizes. Hence the results may have finite size effects, and if $b$ 
really approaches unity for both WSA and WSB in the thermodynamic limit, one can conclude that the finite size 
effects are more  for WSB.

The average clustering coefficient vanishes as $p \to 1$ in both WS types of 
models with
an exponent that is markedly different and can be explained why. The
variance of the degree distribution on the other hand
shows difference in the magnitude only. The clustering coefficients as
functions of degree are qualitatively similar in WSA and WSB models 
although in the latter, the dependence on $k$ is stronger as $p \to 1$.

For the EA and EB models also, we find the qualitative results to
be similar. The presence of the phase transition at $\alpha=2$ is
evident in both models, the shortest paths in EA showing a sharper 
transition. 
Once again, we have estimated the shortest paths exactly which restricts the 
system sizes considerably such that the position of the 
second transition (below $\alpha = 2$), if any,  is difficult to establish numerically. However, analysis of the clustering coefficients 
indicate the presence of another transition at $\alpha_{c}$, the exact value of $\alpha_c$ 
being different in EA and EB. The exponents,
however, are identical and comparable to the model with a lower
average degree studied earlier \cite{biswas}.

One interesting point to note is that apart from the shortest paths, no other
quantity shows any difference for EA and EB beyond $\alpha =2$. 
Although the clustering and degree distribution for EA and EB are identical for $\alpha > 2.0$, there is an interesting behaviour of the clustering coefficient which increases beyond the value 0.5 corresponding to the regular network.

The Euclidean model B has been considered to be appropriate to model
linear polymers or self avoiding walks \cite{psbkc,mouk} . The value of $\alpha$ which will be comparable to a linear polymer  is
less than $2$. However, the absence of magnetic phase transition on a 
polymer chain indicates that it behaves as a regular network while for the Euclidean model
B, the behaviour at $\alpha < 2$ is definitely different from that of a linear chain. This indicates
that the EB does not really mimic the behaviour of a linear polymer possibly because in the
former, the long ranged links are uncorrelated which is not true for a polymer chain where the bridges 
are correlated.  

As mentioned in section II, the WS and Euclidean models are equivalent 
at extreme limiting values of the parameter $\alpha $ and $p$. 
So it also makes sense to compare  the two when 
the parameters assume intermediate values.
The main difference is of course that there is at least one non-trivial phase transition in the Euclidean models. Increasing $\alpha$ beyond 2, we have an extensive region where regular network-like behaviour can be observed. 
No such study is possible for the WS type networks, which 
behaves like a small world network for any nonzero value of $p \neq 1$.
In case one concludes that the Euclidean networks have a random nature below 
$\alpha_c$, then there is a finite region $0 < \alpha <\alpha_{c}$ with
random network behaviour. For the WS type networks, random network 
behaviour exists 
for $p=1$ only.

In both models, the degree distributions have exponential 
tails. The variance in the Euclidean network however shows a 
constant value over a considerable
region, which is not observed in the WS models. This may be due to 
the random network like behaviour of the Euclidean models in the entire region 
$\alpha < \alpha_c$.

The clustering coefficient as function of the parameter $p$ or $\alpha$ shows that it 
has a scaling behaviour as the point $p=1$ or $\alpha = \alpha_{c}$ is approached. 
While for WS type networks, the exponents are appreciably different for the different schemes of rewiring, for the Euclidean networks, the two exponents $\gamma$ and $\delta$ are identical for EA 
and EB while the value of $\alpha_{c}$ shows a difference.

For WSA each node has at least $K/2 = 2$ edges after the rewiring process \cite{bareview}. So there are no isolated
 nodes and the network is usually connected. But for EA there is no such restriction and one gets broken networks for EA with much higher probability.
In our simulations broken WSA networks were not obtained at all (unless $L$ is very small) while 
for the EA, such networks, even for large values of $L$ were generated and were not 
considered for the various calculations.

In summary, we have made a detailed study of different properties of 
some models of networks in which the randomness has been incorporated in 
different ways. We conclude that small changes in the randomness in network models do not 
alter the gross qualitative features.\\

Acknowledgments:  Financial support from DST project SR/S2/CMP-56/2007 and partial computational support from UPE project  
are  acknowledged (SB and PS). SG acknowledges financial support from CSIR ( Grant no. 09/028(0762)/2010-EMR-I).

\end{document}